\documentclass[aps,prd,floatfix,twocolumn,superscriptaddress,preprintnumbers,nofootinbib]{revtex4-1}

\usepackage{textcomp} 
\usepackage{slashed}
\usepackage{epsfig,latexsym,cancel,amssymb,amsmath,verbatim,mathrsfs}
\usepackage{color}
\usepackage{graphicx}
\usepackage{enumitem}
\usepackage[colorlinks,citecolor=blue]{hyperref}

\newcommand{\beq}{\begin{equation}}
\newcommand{\eeq}{\end{equation}}
\newcommand{\bea}{\begin{eqnarray}}
\newcommand{\eea}{\end{eqnarray}}
\newcommand{\nn}{\nonumber}

\allowdisplaybreaks[1]

\begin{document}

\preprint{
	{\vbox {
			\hbox{\bf LA-UR-22-21939}
}}}
\vspace*{0.2cm}

\title{Probing the dark photon via polarized  DIS scattering at the HERA and EIC}

\author{Bin Yan}
\email{yanbin@ihep.ac.cn}
\affiliation{Institute of High Energy Physics, Chinese Academy of Sciences, Beijing 100049, China}
\affiliation{Theoretical Division, Group T-2, MS B283, Los Alamos National Laboratory, P.O. Box 1663, Los Alamos, NM 87545, USA}

\begin{abstract}
The dark photon is widely predicted in many new physics beyond the Standard Model. In this work, we propose to utilize the polarized lepton cross section and single-spin asymmetry in neutral current DIS process at the HERA and upcoming Electron-Ion Collider (EIC) to probe the properties of  the dark photon. It shows that we can constrain the mixing parameter  $\epsilon<0.02$ when the dark photon mass $m_{A_D}<10 ~{\rm GeV}$  at the HERA  with polarized lepton beam and this bound is comparable to the limit from the unpolarized data of HERA I and II. With the help of high integrated luminosity and high electron beam polarization of the EIC, the upper limit of the $\epsilon$ could be further improved. 
Depending on the assumption of the systematic error of the cross section measurements, we obtain $\epsilon<0.01\sim 0.02$ for the mass region $m_{A_D}<10~{\rm GeV}$ under the integrated luminosity of $300~{\rm fb}^{-1}$.  A similar and robust upper limit for the $\epsilon$ could be obtained from the  single-spin asymmetry measurement, which is not depending on the assumption  of the systematic errors.
\end{abstract}

\maketitle

\noindent {\bf Introduction:~}Various astrophysical observations suggest that the necessity of the dark matter (DM) in our universe, and it is widely believed that the DM could couple to the Standard Model (SM) particles weakly through some mediators. A well-know example of the mediator is the dark photon which could be generated from the additional $U(1)$ gauge symmetry. 
Since the theoretical framework  of the dark photon is reasonablely simple and it is  readily accessible in the experimental searches, the dark photon has received much attention in the high energy physics community in recent years; e.g. see the recent discussions in Refs.~\cite{An:2012ue,An:2013yua,Curtin:2014cca,An:2014twa,He:2017ord,Gu:2017gle,He:2017zzr,Lindner:2018kjo,Pan:2018dmu,Ilten:2018crw,Fuyuto:2019vfe,Du:2019mlc,Chen:2020bok,An:2020jmf,An:2020bxd,Kribs:2020vyk,Fabbrichesi:2020wbt,Cheng:2021qbl,Dev:2021otb,Thomas:2021lub,Graham:2021ggy,Inan:2021dir,Du:2021cmt,Barducci:2021egn,Su:2021jvk,Caputo:2021eaa,Wong:2021lgk,Thomas:2022qhj,Hosseini:2022urq}.

The dark photon could couple to the SM particles through the kinetic mixing with the $U(1)_Y$ gauge field strength tensor ($B_{\mu\nu}$)~\cite{Holdom:1985ag},
\beq
\mathcal{L}_{\rm eff}=-\frac{1}{4}F_{\mu\nu}^\prime F^{\prime \mu\nu}+\frac{m_{A^\prime}^2}{2}A_\mu^\prime A^{\prime \mu}+\frac{\epsilon}{2c_W}F_{\mu\nu}^\prime B^{\mu\nu},
\label{eq:Lag}
\eeq
where $F_{\mu\nu}^\prime=\partial_\mu A_\nu^\prime-\partial_\nu A_\mu^\prime$ is the field strength tensor of the dark photon, $c_W\equiv \cos\theta_W$ is the cosine of the weak mixing angle $\theta_W$ and the parameter $\epsilon$ denotes the mixing strength between the dark photon and $U(1)_Y$ gauge field.  The search of the dark photon depends on the production mechanisms and decay modes at colliders and at fixed-target or beam dumps. For example, the dark photon could be probed by the meson decays in NA48/2~\cite{NA482:2015wmo}, bremsstrahlung in A1~\cite{Merkel:2014avp}, annihilation in Babar~\cite{BaBar:2014zli}, and all these processes at KLOE~\cite{KLOE-2:2011hhj,KLOE-2:2012lii,KLOE-2:2014qxg,KLOE-2:2016ydq}, the processes at the LHC~\cite{LHCb:2019vmc}.  It shows that the  parameter space of
the light dark photon (below 1 GeV) has been seriously constrained by the experimental data~\cite{Ilten:2018crw,Fabbrichesi:2020wbt}. In the mass region of tens of GeV, the parameter space is less limited and dominantly determined by the LHC measurements~\cite{Ilten:2018crw,Fabbrichesi:2020wbt}.  It shows that the upper limit of the mixing parameter  has been constrained to be around $\epsilon\sim \mathcal{O}(10^{-3}-10^{-2})$. 

However, those bounds are strongly depending on the assumption of the detail structure of the dark sector, due to the dark matter could change the decay branching ratios of the dark photon.  Recently, a ``decay-agnostic" scenario received much attention in probing the dark photon due to the limits are independent of the unknown dark sector.  One of the "decay-agnostic" scenario in probing the dark photon is from the electroweak precision observables (EWPO) of the $Z$-pole at the LEP~\cite{Hook:2010tw,Curtin:2014cca}. The mixing effects between the dark photon and $B{\mu\nu}$ will modify the gauge couplings of the $Z$-boson to SM fermions, as a result, we could learn the information of the dark photon indirectly from those precision measurements. It shows that $\epsilon<0.025$ could be obtained when the light dark photon is far away from the $Z$-pole ($m_{A_D}\ll m_Z$). Recently, a new approach from the unpolarized deeply-inelastic scattering (DIS)  at the HERA has been proposed to further improve the limits of the parameter space of the dark photon~\cite{Kribs:2020vyk}.  After that, a global analysis of combing the parton distribution functions (PDFs) and dark photon at the same time with the unpolarized DIS data at the HERA  has been discussed in Ref.~\cite{Thomas:2021lub}. Furthermore, a single-spin asymmetry from elastic and deeply-inelastic scattering  with a polarized electron beam could also offer an opportunity to probe the dark photon~\cite{Thomas:2022qhj}. In this work, we try to utilize the polarized DIS scattering at the HERA and the upcoming Electron-Ion Collider (EIC) to further probe the properties of the dark photon. We will demonstrate below that the interactions of the dark photon to SM fermions show a chirality structure due to the mixing between the dark photon and $Z$-boson, as a consequence, a properly polarization of the incoming lepton beam could enhance the sensitivity in probing the dark photon.

\vspace{3mm}
\noindent {\bf Dark photon model:~}
To obtain the couplings of the SM particles to the physical dark photon $A_D$ and SM $Z$-boson, we need to diagonalize the mixing term in Eq.~\eqref{eq:Lag} through the field redefinition and normalization; see the detail in Ref.~\cite{Kribs:2020vyk}.  We parameterize the effective couplings of $A_{D}$ and $Z$ boson to the fermion $f$ by,
\begin{align}
\mathcal{L}_{Zf\bar{f}}&=\frac{g_W}{2c_W}\bar{f}\gamma_\mu(\widetilde{g}_V^f-\widetilde{g}_A^f\gamma_5)fZ^\mu,\nn\\
\mathcal{L}_{A_Df\bar{f}}&=\frac{g_W}{2c_W}\bar{f}\gamma_\mu(\widetilde{g}_V^{\prime f}-\widetilde{g}_A^{\prime f}\gamma_5)fA_D^\mu,
\label{eq:Lag}
\end{align}
where $g_W$ is the $SU(2)_L$ gauge coupling.  The vector and axial-vector couplings in Eq.~\eqref{eq:Lag} can be written as~\cite{Kribs:2020vyk},
\begin{align}
\widetilde{g}_V^f&=g_V^f(c_\alpha-\epsilon_W s_\alpha)+2c_W^2\epsilon_W s_\alpha Q_f,\label{eq:geff1}\\
\widetilde{g}_A^f&=g_A^f(c_\alpha-\epsilon_W s_\alpha),\\
\widetilde{g}_V^{\prime f}&=-g_V^f(s_\alpha+\epsilon_W c_\alpha)+2c_W^2\epsilon_W c_\alpha Q_f,\\
\widetilde{g}_A^{\prime f}&=-g_A^f(s_\alpha+\epsilon_W c_\alpha).
\label{eq:geff2}
\end{align}
Here $g_{V,A}^f$ are the vector and axial vector couplings of $Z$-boson to fermion $f$ in the SM, and it shows that 
$g_V^f=T_3^f-2Q_fs_W^2$ and $g_A^f=T_3^f$, where $(T_3^f,Q_f)=(1/2,2/3), (-1/2,-1/3)$ and $(-1/2,-1)$ for up-type, down-type quark and electron, respectively, with $s_W\equiv\sin\theta_W$. The cosine and sine of the mixing angle $\alpha$ denote as $c_\alpha$ and $s_\alpha$, respectively and can be calculated by,
\begin{align}
\tan\alpha&=\frac{1}{2\epsilon_W}\left[1-\epsilon_W^2-\rho^2\right.\nn\\
&\left.-{\rm sign}(1-\rho^2)\sqrt{4\epsilon_W^2+\left(1-\epsilon_W^2-\rho^2\right)^2}\right],
\label{eq:mix}
\end{align}
with
\begin{align}
\epsilon_W&=\frac{\epsilon \tan\theta_W}{\sqrt{1-\epsilon^2/c_W^2}},&
\rho&=\frac{m_{A^\prime}/m_Z^0}{\sqrt{1-\epsilon^2/c_W^2}}.
\end{align}
Here $m_Z^0$ is the $Z$-boson mass before the mixing. The masses of the physical $Z$-boson and dark photon are given by,
\begin{align}
m_{Z,A_D}^2&=\frac{(m_Z^0)^2}{2}\left[1+\epsilon_W^2+\rho^2\right.\nn\\
&\left.\pm{\rm sign}(1-\rho^2)\sqrt{\left(1+\epsilon_W^2+\rho^2\right)^2-4\rho^2}\right].
\end{align}
The mixing angle $\alpha$ in Eq.~\eqref{eq:mix} can be translated as a function of parameters $(m_Z,m_{A_D},\epsilon)$. In the limit of $\epsilon\to 0$, we have $m_Z=m_Z^0$, $m_{A_D}=m_{A^\prime}$ and $\alpha,\epsilon_W\sim\mathcal{O}(\epsilon)$.  Therefore, the correction of the dark photon to the Z boson couplings arises from $\mathcal{O}(\epsilon^2)$, while the gauge couplings of the dark photon are $\mathcal{O}(\epsilon)$; see Eqs.~\eqref{eq:geff1} -\eqref{eq:geff2}. 

\vspace{3mm}
\noindent {\bf DIS cross section:~}
In this section, we review the neutral current DIS cross section and refer the reader to, e.g.~\cite{Blumlein:2012bf,Zyla:2020zbs,Cirigliano:2021img} for the detail of the factorization of the DIS production. For the process $e^-(k)+p(P)\to e^-(k^\prime)+X$ with a polarized incoming electron, we can write the cross section in terms of the structure functions~\cite{Zyla:2020zbs},
\beq
\frac{d\sigma}{\sigma_0dxdy}=F_1\left((1-y)^2+1\right)+F_L\frac{1-y}{x}+F_3\lambda_e\left(y-\frac{y^2}{2}\right),
\label{eq:DIS}
\eeq
where $\lambda_e=\pm 1$ is the helicity of the incoming electron,  $\sigma_0\equiv 4\pi\alpha_{\rm em}^2/(xy^2S)$ with $S=(k+P)^2$ and $\alpha_{\rm em}$ is the fine-structure constant. Note that the structure functions $F_{1,L}\equiv F_{1,L}(x,y)$ are related to the parity conserved interactions in hadronic tensor, while $F_3\equiv F_3(x,y)$ is corresponding to the parity-violating scattering of quarks.  The $F_L$ is usually defined as the combination of the structure functions $F_{1,2}$,  i.e. $F_L=F_2-xF_1$.  The  DIS kinematic variables are defined as,
\beq
Q^2=-q^2,\quad x=\frac{Q^2}{2P\cdot q},\quad y=\frac{P\cdot q}{P\cdot k}, \quad xyS=Q^2.
\eeq
Here $q_\mu=k_\mu-k_\mu^\prime$ denotes the momentum transfer of the electrons. The structure functions $F_i$ with $i=1,2,3$ in this work contain the contributions originated from photon-only ($F_i^\gamma$), $Z$-only ($F_i^Z$), $\gamma-Z$ interference ($F_i^{\gamma Z}$), $\gamma-A_D$ interference ($F_i^{\gamma A_D}$), $A_D$ only ($F_i^{A_D}$) and $A_D-Z$ interference channels ($F_i^{A_D Z}$), and can be written as
\begin{align}
F_i&=F_i^\gamma-(\widetilde{g}_V^e-\lambda_e\widetilde{g}_A^e)\eta_{\gamma Z}F_i^{\gamma Z}+(\widetilde{g}_V^e-\lambda_e\widetilde{g}_A^e)^2\eta_{Z}F_i^{Z}\nn\\
&-(\widetilde{g}_V^{\prime e}-\lambda_e\widetilde{g}_A^{\prime e})\eta_{\gamma A_D}F_i^{\gamma A_D}+(\widetilde{g}_V^{\prime e}-\lambda_e\widetilde{g}_A^{\prime e})^2\eta_{A_D}F_i^{A_D}\nn\\
&+(\widetilde{g}_V^e-\lambda_e\widetilde{g}_A^e)(\widetilde{g}_V^{\prime e}-\lambda_e\widetilde{g}_A^{\prime e})\eta_{A_DZ}F_i^{A_DZ},
\label{eq:FF}
\end{align}
where $\eta_i$ denotes the ratio of the gauge coupling and propagator from massive dark photon or $Z$-boson to the photon's, and it shows
\begin{align}
\eta_{\gamma Z}&=\frac{Q^2}{Q^2+m_Z^2}\frac{1}{4c_W^2s_W^2},& \eta_Z&=\eta_{\gamma Z}^2,\nn\\
\eta_{\gamma A_D}&=\frac{Q^2}{Q^2+m_{A_D}^2}\frac{1}{4c_W^2s_W^2},& \eta_{A_D}&=\eta_{\gamma A_D}^2,\nn\\
\eta_{A_DZ}&=\frac{Q^4}{(Q^2+m_Z^2)(Q^2+m_{A_D}^2)}\frac{1}{16c_W^4s_W^4}.
\end{align}
For $m_{A_D}\ll m_Z$, the leading effects of the dark photon to the DIS cross section is from the interference between photon and $A_D$, i.e. $F_i^{\gamma A_D}$ in Eq.~\eqref{eq:FF}. The other terms will be suppressed by $Z$ propagator or more power of the mixing parameter $\epsilon$ if $\epsilon\ll 1$. We also notice that the gauge couplings $\widetilde{g}_V^{\prime e}\simeq -1.48\epsilon$ and  $\widetilde{g}_A^{\prime e}\simeq 0.77\epsilon$ in the limit $\epsilon\ll 1$, as a consequence, the contribution from dark photon will be enhanced for $\lambda_e=1$, while it will be suppressed for $\lambda_e=-1$; see Eq.~\eqref{eq:FF}. Therefore, we could expect that  a right-handed electron beam in DIS scattering could give a stronger limit for the dark photon than the left-handed and unpolarized electron beams. On the other hand, when $m_{A_D}\sim m_Z$, both the interference between $Z/A_D$ and $\gamma$ could give a comparable contribution to the DIS cross section. The structure functions in the SM with massless and massive quarks have been extensive discussed in the literature~\cite{vanNeerven:1991nn,Zijlstra:1992qd,Zijlstra:1992kj,Larin:1996wd,Moch:1999eb,Bierenbaum:2007qe,Bierenbaum:2009mv,Guzzi:2011ew,Kawamura:2012cr}. In this work, we follow the discussion in Ref.~\cite{Kang:2014qba} to calculate the structure functions $F_i$ at the next-to-leading order (NLO) in strong coupling $\alpha_s$.

\vspace{3mm}
\noindent {\bf Sensitivity at the HERA:~}
Both  the electron and positron beams,  with different degrees of polarization and luminosities have been used in HERA experiments; see Table~\ref{tab:lumi} for the detail information of the data sets. The cross section for the positron beam can be obtained from Eq.~\eqref{eq:DIS} with the replacement of $\lambda_e\to-\lambda_e$, due to the opposite helicity between the particle and anti-particle. The cross section from the experimental measurement with lepton beam polarization $P_e$ can be written as a combination of the incoming lepton at its helicity eigenstates,
\beq
\sigma(P_e)=\frac{1}{2}(\sigma_++\sigma_-)+\frac{P_e}{2}(\sigma_+-\sigma_-),
\eeq
where $\sigma_\pm$ is the inclusive DIS cross section of a right-handed $(+)$ or left-handed $(-)$ lepton beam scattering off an unpolarized proton beam.

\setlength{\tabcolsep}{8pt}
\begin{table}
	\begin{center}
		\begin{tabular}{r|c|c}
			\hline
			H1& $R$ & $L$\\
			\hline
			$e^-p$ 
			& $47.3\,{\rm pb}^{-1}$, $0.36$  & $104.4\,{\rm pb}^{-1}$, $-0.258$  \\
			\hline
			$e^+p$
			& $101.3\,{\rm pb}^{-1}$, $0.325$ & $80.7\,{\rm pb}^{-1}$, $-0.37$  \\
			\hline
			\hline
			ZEUS& $R$ & $L$\\
			\hline
			$e^-p$
			& $71.2\,{\rm pb}^{-1}$, $0.29$ & $98.7\,{\rm pb}^{-1}$, $-0.27$ \\
			\hline
			$e^+p$  
			& $78.8\,{\rm pb}^{-1}$, $0.32$ & $56.7\,{\rm pb}^{-1}$, $-0.36$ \\
			\hline
		\end{tabular} 
		\caption{The integrated luminosity and lepton beam's longitudinal polarization for each data set of H1~\cite{H1:2012qti} and ZEUS~\cite{ZEUS:2009swh,ZEUS:2012zcp}. $R$ ($L$) denotes right-handed (left-handed) lepton data set.}
		\label{tab:lumi}
	\end{center}
\end{table}

In HERA experiments, the reduced cross section is widely used, which is defined as~\cite{H1:2012qti,ZEUS:2009swh,ZEUS:2012zcp},
\beq
\widetilde{\sigma}(x,Q^2)=\frac{d\sigma}{dxdQ^2}\frac{xQ^4}{2\pi\alpha^2_{\rm em}}\frac{1}{1+(1-y)^2}.
\eeq
To constrain the dark photon in the parameter space $(\epsilon,m_{A_D})$, we calculate the reduced cross sections based on the $(Q^2,x)$ grids in H1~\cite{H1:2012qti} and ZEUS~\cite{ZEUS:2009swh,ZEUS:2012zcp} collaborations with collider energy $E_{\rm cm}=318~{\rm GeV}$ and the polarization of beams. The double differential distribution $d\sigma/dxdQ^2$ has been calculated based on the CT18 NLO PDFs~\cite{Hou:2019efy}  at the NLO accuracy in $\alpha_s$. Both the renormalization and factorization scales are chosen as $\mu=Q$. To combine the polarized data sets from the HERA analysis, we define the $\chi^2$ as,
\beq
\chi^2=\sum_i\left[\frac{\widetilde{\sigma}_i^{\rm th}-\widetilde{\sigma}_i^{\rm exp}}{\delta \widetilde{\sigma}_i}\right]^2,
\label{eq:chi2}
\eeq
where $\widetilde{\sigma}_i^{\rm th}$ and $\widetilde{\sigma}_i^{\rm exp}$ are, respectively, the theoretical predictions at the NLO accuracy and the experimental values for the $i$-th data ( the reduced cross sections for  the given grids $(Q^2,x)$ from H1 and ZEUS collaboration measurements, see Tables 13-16 in Ref.~\cite{H1:2012qti}  for H1 collaboration, Tables 7-8 in Ref.~\cite{ZEUS:2012zcp} for positron beam and Tables 11,13  in Ref.~\cite{ZEUS:2009swh} for electron beam of the ZEUS collaboration). $\delta \widetilde{\sigma}_i$ denotes the total error of the $i$-th data. 

\begin{figure}
\centering
\includegraphics[scale=0.33]{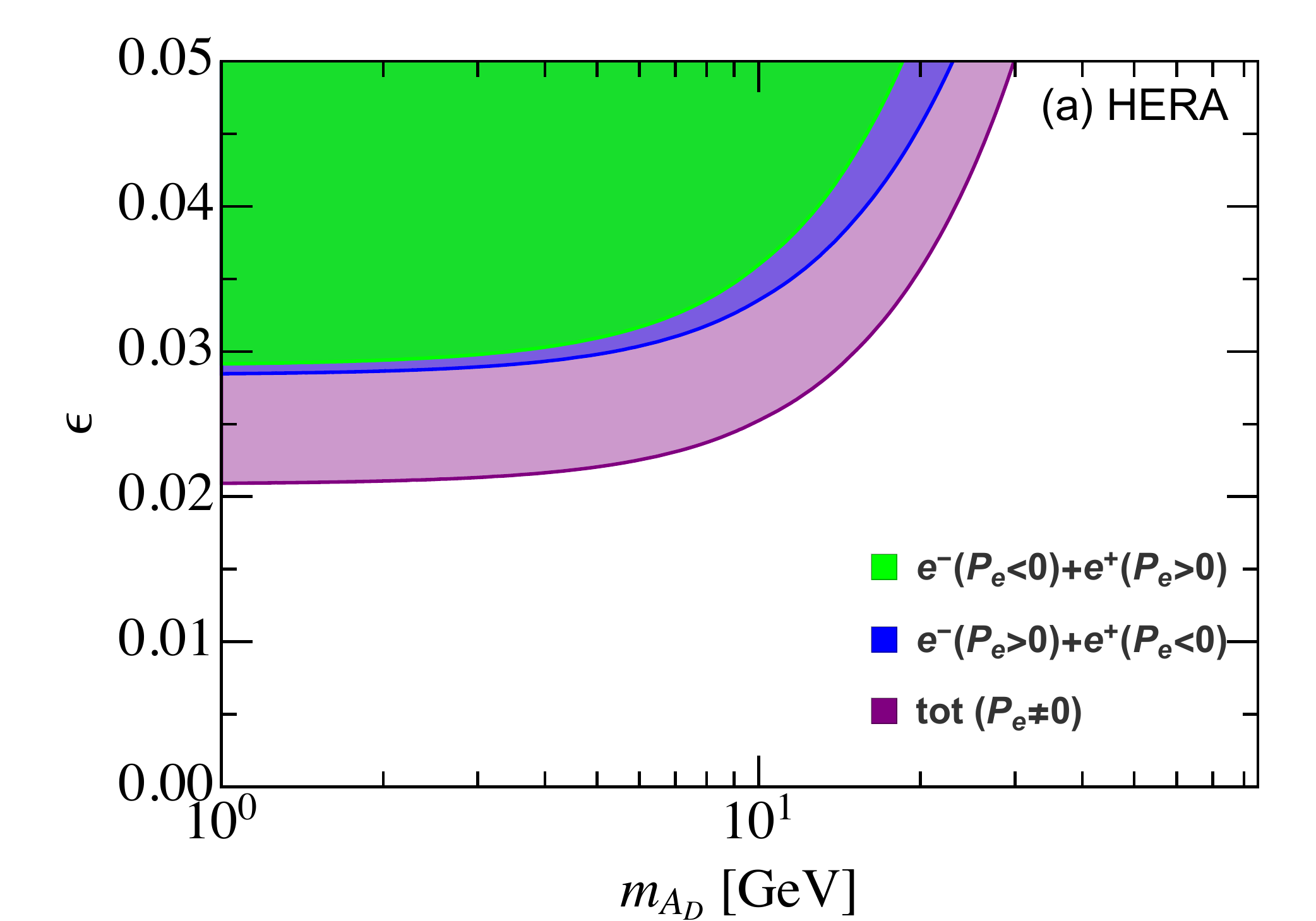}
\includegraphics[scale=0.33]{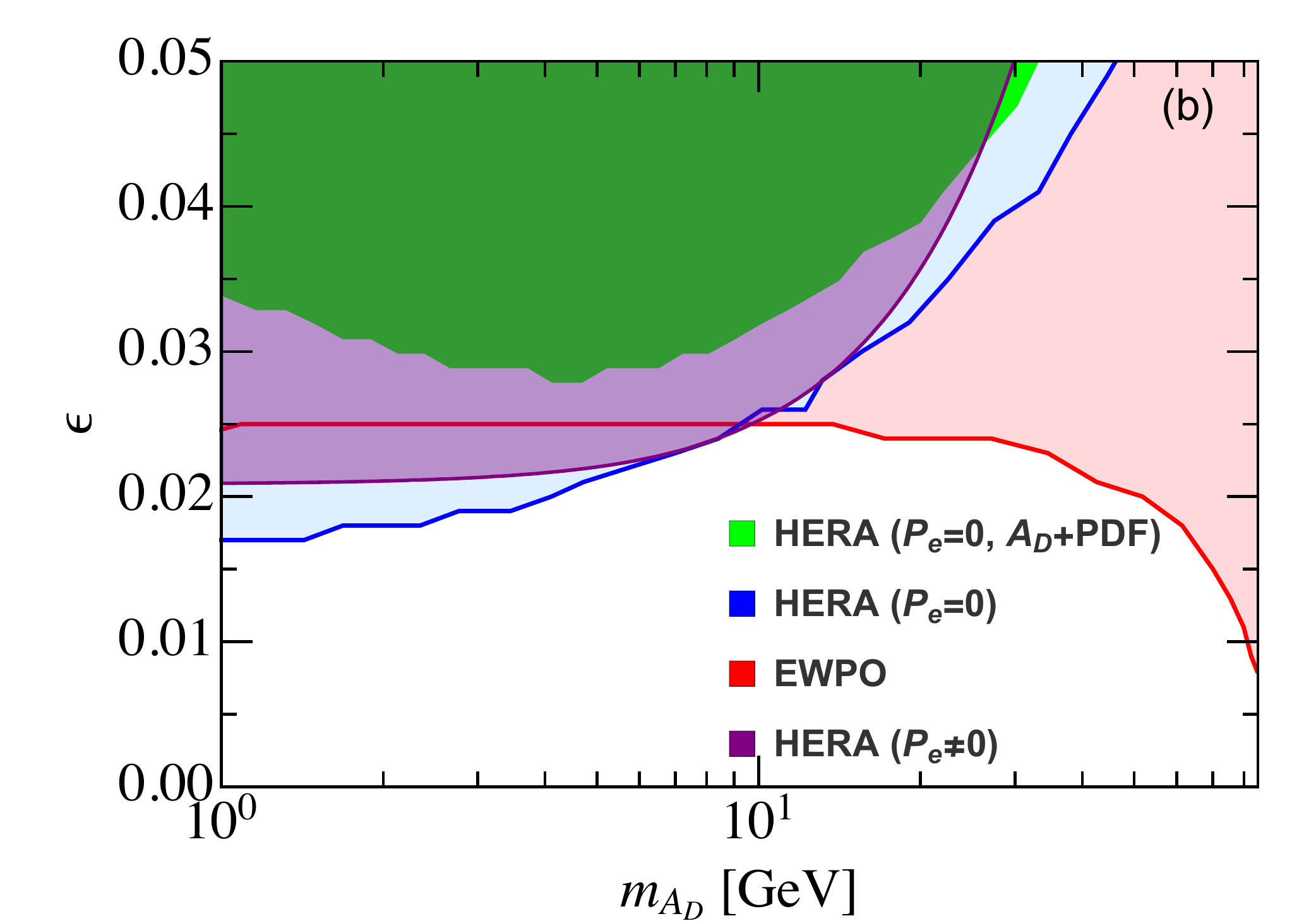}
\caption{The upper limit on the mixing parameter $\epsilon$ from the polarized HERA data at the 95\% C.L.. The excluded regions from unpolarized HERA data and EWPO measurements are extracted from Refs.~\cite{Kribs:2020vyk,Thomas:2021lub}.}
\label{Fig:HERA}
\end{figure}

In Fig.~\ref{Fig:HERA}(a), we show the upper limit of the mixing parameter $\epsilon$ from the various polarized HERA data at the 95\% confidence level (C.L.). 
The green band denotes the limit from $e^-(P_e<0)$ and $e^+(P_e>0)$ beams, while the blue band is corresponding to the constraint from $e^-(P_e>0)$ and $e^+(P_e<0)$. The purple region is the result after including all the polarized data at the HERA.
Although the integrated luminosity from $e^-(P_e>0)$ and $e^+(P_e<0)$ beams are smaller than $e^-(P_e<0)$ and $e^+(P_e>0)$ (see table~\ref{tab:lumi}), the reduced cross sections from $e^-(P_e>0)$ and $e^+(P_e<0)$ beams can give a stronger limit on the mixing parameter $\epsilon$; see Fig.~\ref{Fig:HERA}(a).  It is owing to the fact that the dark photon could sizably change the DIS cross section for the right-handed (left-handed) electron (positron) beams; see Eq.~\eqref{eq:FF}. For the purpose of the comparsion, we show the bounds from the unpolarized HERA data and EWPO measurements in Fig.~\ref{Fig:HERA}(b).  The green and blue bands are corresponding to the global fitting results with and without considering PDFs~\cite{Kribs:2020vyk,Thomas:2021lub}.
It is evident from the purple and blue bands of Fig.~\ref{Fig:HERA}(b) that the limits for the dark photon from the  polarized data of HERA II is comparable to the results of the unpolarized HERA data with a much smaller integrated luminosity.

Owing to couplings of dark photon to the SM fermions violate the parity, it would be convenient to define a single-spin asymmetry (SSA) to directly measure the electroweak effects of dark photon~\cite{Thomas:2022qhj}. Recently, this observable has been widely used to constrain the $Zb\bar{b}$ anomalous couplings at the HERA and EIC~\cite{Yan:2021htf,Li:2021uww}. The definition of the SSA in H1 and ZEUS collaborations are different. For the H1 collaboration, it is~\cite{H1:2012qti},
\beq
A_{H1}=\frac{2}{P_L-P_R}\frac{\sigma(P_L)-\sigma(P_R)}{\sigma(P_L)+\sigma(P_R)},
\eeq
where $P_{L,R}$ is the longitudinal lepton beam polarization in the left- and right-handed data sets, respectively; see Table~\ref{tab:lumi}. The SSA in ZEUS collaboration is defined in terms of $\sigma_{\pm}$~\cite{ZEUS:2009swh,ZEUS:2012zcp}, i.e.,
\beq
A_{\rm ZEUS}=\frac{\sigma_+-\sigma_-}{\sigma_++\sigma_-}.
\label{eq:SSAZ}
\eeq
We can translate the SSA in Eq.~\eqref{eq:SSAZ} to the experimental measurement by
\beq
A_{\rm ZEUS}=\frac{\sigma(P_R)-\sigma(P_L)}{P_R\sigma(P_L)-P_L\sigma(P_R)}.
\eeq
Those two definitions are equivalent to each other when $P_L=-P_R$. Similar to the $\chi^2$ analysis in reduced cross sections, we can also combine all the SSA measurements at the HERA to constrain the parameter space of dark photon. However, it shows that the band of the dark photon can not be improved by the SSA measurements due to the large experimental uncertainties.

\vspace{3mm}
\noindent {\bf Sensitivity at the EIC:~}
In this section, we explore the potential of probing the dark photon at the upcoming EIC.  Although the expected collider energy $E_{\rm cm}=141~{\rm GeV}$ at the EIC is lower than the HERA, the integrated luminosity and lepton beam polarization could be very large~\cite{AbdulKhalek:2021gbh}. These advantages at the EIC offer the possibility to improve the dark photon measurement.  To simplify the analysis, we only consider the electron beam and assume the degree of polarization and integrated luminosities for right-handed and left-handed electron beams are the same.  Moreover, 
we consider the polarized DIS cross sections instead of the reduced cross sections in our analysis. 
To avoid the non-perturbative QCD effects, we focus on the following EIC kinematic region: $x\in[0.005,0.8]$ and $Q^2\in [10^2,10^4]~{\rm GeV}^2$~\cite{AbdulKhalek:2021gbh}. We consider the DIS cross sections in separate bins in $(Q^2,x)$ space by assuming the statistical errors are of the same order as the systematic error.

\begin{figure}
\centering
\includegraphics[scale=0.33]{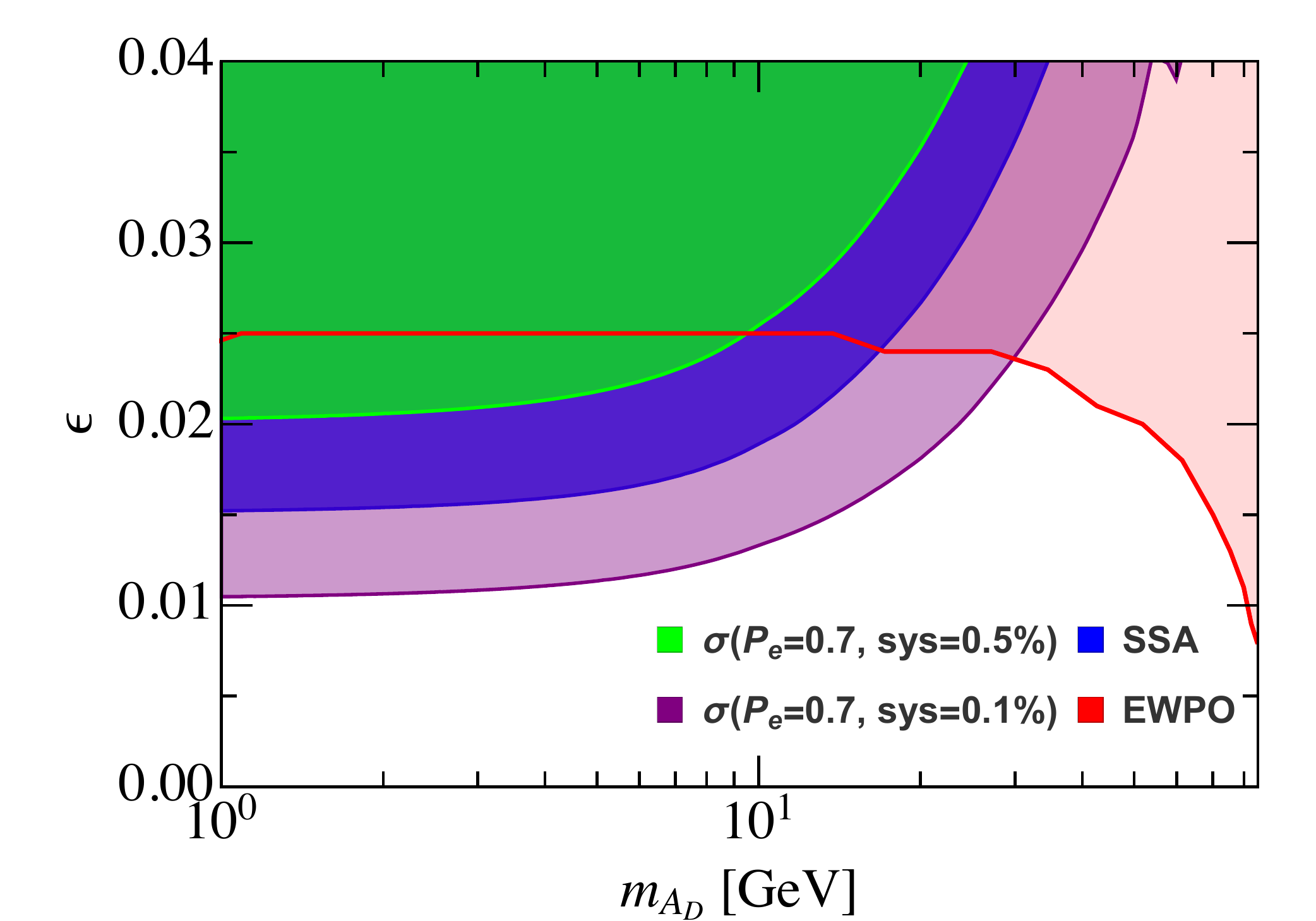}
\caption{The expected sensitivity on the mixing parameter $\epsilon$ from the polarized EIC data at the 95\% C.L. The integrated luminosity for the polarized DIS cross sections is assumed to be $300~{\rm fb}^{-1}$, while both the right- and left-handed electron beams are $150~{\rm fb}^{-1}$  for the SSA measurement. }
\label{Fig:EIC}
\end{figure}

We perform a $\chi^2$ analysis as Eq.~\eqref{eq:chi2}  to constrain the parameter space of the dark photon.  The experimental values of the cross sections are assumed to 
be the same as the SM predictions and the electron beam polarization $P_e\equiv P_R=-P_L=0.7$~\cite{AbdulKhalek:2021gbh} will be used in our calculation. Figure~\ref{Fig:EIC} shows the expected sensitivity of the mixing parameter $\epsilon$ from the polarized EIC data at the 95\% C.L. The integrated luminosity is assumed to be $300~{\rm fb}^{-1}$ for the polarized cross sections. The bounds from the left-handed ($P_e=-0.7$) and unpolarized ($P_e=0$) electron beams with $300~{\rm fb}^{-1}$ would be weaker than right-handed ($P_e=0.7$) electron data set, as a result, we only show the expected limit from the right-handed electron data in Fig.~\ref{Fig:EIC}. We consider two systematic errors in each bin in our study, i.e. 0.1\% (green band) and 0.5\% (purple band).  Depending on the assumption of  the systematic errors in cross section measurements, we could obtain a model-independent bound on the $\epsilon<0.01\sim 0.02$ for $m_{A_D}<10~{\rm GeV}$ at the EIC. We also notice that the purple band in Fig.~\ref{Fig:EIC} is dominantly determined by the systematic error. In this same figure, we also consider the limit from the SSA (blue band), which was defined by
\beq
A_{\rm EIC}=\frac{1}{P_e}\frac{\sigma(P_e)-\sigma(-P_e)}{\sigma(P_e)+\sigma(-P_e)}.
\eeq
The advantage of the SSA is that the systematic uncertainties could be cancelled through the definition and can be ignored in this work~\cite{ZEUS:2009swh}. The statistical uncertainty of SSA is given by
\beq
\delta A_{\rm EIC}\simeq\frac{1/P_e}{\sqrt{2\mathcal{L}\sigma(P_e=0)}}.
\eeq
where $\mathcal{L}$ is the integrated luminosity of the right- and left-handed electron data sets. From the blue band of Fig.~\ref{Fig:EIC}, we can obtain the upper limit of the mixing parameter  $\epsilon<0.015$ when $m_{A_D}<10~{\rm GeV}$. We emphasize that this conclusion is not sensitive to the assumption of the systematic error of the measurement.
Before closing this section, we would like to comment the possible limits from the low energy polarized electron scattering~\cite{Arcadi:2019uif} and neutrino-electron scattering~\cite{Bilmis:2015lja}.  We note that these measurements would be more sensitivity to the very light dark photon,  while the limits for the parameter space of we are considering should be very weak.

\vspace{3mm}
\noindent {\bf Conclusions:~}
In this work, we suggest to utilize the DIS cross section and single-spin asymmetry with polarized lepton beam at the HERA and EIC to probe the properties of the dark photon. Owing to the mixing between the dark photon and $Z$ boson, the gauge couplings of dark photon violate the parity, as a result, we find that the polarized DIS cross section from right-handed electron beam could give a stronger limit on the parameter space of the dark photon than the left-handed and unpolarized lepton beams. We further demonstrate that the limit from the polarized HERA data is comparable to the bound from the unpolarized measurements. With enough integrated luminosity collected at the EIC, it is possible to further improve the upper limit of the mixing parameter by the polarized cross section and single-spin asymmetry measurements.  We emphasize that the sensitivity from SSA is not depending on the assumption of the systematic error, but it is not for the cross section measurements.

\medskip
\noindent{\bf Acknowledgments.}
The author thanks the helpful discussion of C.-P. Yuan. This work is supported by the U.S. Department of Energy, Office of Science,
Office of Nuclear Physics, under Contract DE-AC52-06NA25396 through the LANL/LDRD Program, as well as the TMD topical collaboration for nuclear theory.

\bibliographystyle{apsrev}
\bibliography{reference}

\begin{thebibliography}{61}
\expandafter\ifx\csname natexlab\endcsname\relax\def\natexlab#1{#1}\fi
\expandafter\ifx\csname bibnamefont\endcsname\relax
  \def\bibnamefont#1{#1}\fi
\expandafter\ifx\csname bibfnamefont\endcsname\relax
  \def\bibfnamefont#1{#1}\fi
\expandafter\ifx\csname citenamefont\endcsname\relax
  \def\citenamefont#1{#1}\fi
\expandafter\ifx\csname url\endcsname\relax
  \def\url#1{\texttt{#1}}\fi
\expandafter\ifx\csname urlprefix\endcsname\relax\def\urlprefix{URL }\fi
\providecommand{\bibinfo}[2]{#2}
\providecommand{\eprint}[2][]{\url{#2}}

\bibitem[{\citenamefont{An et~al.}(2013{\natexlab{a}})\citenamefont{An, Huo,
  and Wang}}]{An:2012ue}
\bibinfo{author}{\bibfnamefont{H.}~\bibnamefont{An}},
  \bibinfo{author}{\bibfnamefont{R.}~\bibnamefont{Huo}}, \bibnamefont{and}
  \bibinfo{author}{\bibfnamefont{L.-T.} \bibnamefont{Wang}},
  \bibinfo{journal}{Phys. Dark Univ.} \textbf{\bibinfo{volume}{2}},
  \bibinfo{pages}{50} (\bibinfo{year}{2013}{\natexlab{a}}), \eprint{1212.2221}.

\bibitem[{\citenamefont{An et~al.}(2013{\natexlab{b}})\citenamefont{An,
  Pospelov, and Pradler}}]{An:2013yua}
\bibinfo{author}{\bibfnamefont{H.}~\bibnamefont{An}},
  \bibinfo{author}{\bibfnamefont{M.}~\bibnamefont{Pospelov}}, \bibnamefont{and}
  \bibinfo{author}{\bibfnamefont{J.}~\bibnamefont{Pradler}},
  \bibinfo{journal}{Phys. Rev. Lett.} \textbf{\bibinfo{volume}{111}},
  \bibinfo{pages}{041302} (\bibinfo{year}{2013}{\natexlab{b}}),
  \eprint{1304.3461}.

\bibitem[{\citenamefont{Curtin et~al.}(2015)\citenamefont{Curtin, Essig, Gori,
  and Shelton}}]{Curtin:2014cca}
\bibinfo{author}{\bibfnamefont{D.}~\bibnamefont{Curtin}},
  \bibinfo{author}{\bibfnamefont{R.}~\bibnamefont{Essig}},
  \bibinfo{author}{\bibfnamefont{S.}~\bibnamefont{Gori}}, \bibnamefont{and}
  \bibinfo{author}{\bibfnamefont{J.}~\bibnamefont{Shelton}},
  \bibinfo{journal}{JHEP} \textbf{\bibinfo{volume}{02}}, \bibinfo{pages}{157}
  (\bibinfo{year}{2015}), \eprint{1412.0018}.

\bibitem[{\citenamefont{An et~al.}(2015)\citenamefont{An, Pospelov, Pradler,
  and Ritz}}]{An:2014twa}
\bibinfo{author}{\bibfnamefont{H.}~\bibnamefont{An}},
  \bibinfo{author}{\bibfnamefont{M.}~\bibnamefont{Pospelov}},
  \bibinfo{author}{\bibfnamefont{J.}~\bibnamefont{Pradler}}, \bibnamefont{and}
  \bibinfo{author}{\bibfnamefont{A.}~\bibnamefont{Ritz}},
  \bibinfo{journal}{Phys. Lett. B} \textbf{\bibinfo{volume}{747}},
  \bibinfo{pages}{331} (\bibinfo{year}{2015}), \eprint{1412.8378}.

\bibitem[{\citenamefont{He et~al.}(2017)\citenamefont{He, He, and
  Huang}}]{He:2017ord}
\bibinfo{author}{\bibfnamefont{M.}~\bibnamefont{He}},
  \bibinfo{author}{\bibfnamefont{X.-G.} \bibnamefont{He}}, \bibnamefont{and}
  \bibinfo{author}{\bibfnamefont{C.-K.} \bibnamefont{Huang}},
  \bibinfo{journal}{Int. J. Mod. Phys. A} \textbf{\bibinfo{volume}{32}},
  \bibinfo{pages}{1750138} (\bibinfo{year}{2017}), \eprint{1701.08614}.

\bibitem[{\citenamefont{Gu and He}(2018)}]{Gu:2017gle}
\bibinfo{author}{\bibfnamefont{P.-H.} \bibnamefont{Gu}} \bibnamefont{and}
  \bibinfo{author}{\bibfnamefont{X.-G.} \bibnamefont{He}},
  \bibinfo{journal}{Phys. Lett. B} \textbf{\bibinfo{volume}{778}},
  \bibinfo{pages}{292} (\bibinfo{year}{2018}), \eprint{1711.11000}.

\bibitem[{\citenamefont{He et~al.}(2018)\citenamefont{He, He, Huang, and
  Li}}]{He:2017zzr}
\bibinfo{author}{\bibfnamefont{M.}~\bibnamefont{He}},
  \bibinfo{author}{\bibfnamefont{X.-G.} \bibnamefont{He}},
  \bibinfo{author}{\bibfnamefont{C.-K.} \bibnamefont{Huang}}, \bibnamefont{and}
  \bibinfo{author}{\bibfnamefont{G.}~\bibnamefont{Li}}, \bibinfo{journal}{JHEP}
  \textbf{\bibinfo{volume}{03}}, \bibinfo{pages}{139} (\bibinfo{year}{2018}),
  \eprint{1712.09095}.

\bibitem[{\citenamefont{Lindner et~al.}(2018)\citenamefont{Lindner, Queiroz,
  Rodejohann, and Xu}}]{Lindner:2018kjo}
\bibinfo{author}{\bibfnamefont{M.}~\bibnamefont{Lindner}},
  \bibinfo{author}{\bibfnamefont{F.~S.} \bibnamefont{Queiroz}},
  \bibinfo{author}{\bibfnamefont{W.}~\bibnamefont{Rodejohann}},
  \bibnamefont{and} \bibinfo{author}{\bibfnamefont{X.-J.} \bibnamefont{Xu}},
  \bibinfo{journal}{JHEP} \textbf{\bibinfo{volume}{05}}, \bibinfo{pages}{098}
  (\bibinfo{year}{2018}), \eprint{1803.00060}.

\bibitem[{\citenamefont{Pan et~al.}(2020)\citenamefont{Pan, He, He, and
  Li}}]{Pan:2018dmu}
\bibinfo{author}{\bibfnamefont{J.-X.} \bibnamefont{Pan}},
  \bibinfo{author}{\bibfnamefont{M.}~\bibnamefont{He}},
  \bibinfo{author}{\bibfnamefont{X.-G.} \bibnamefont{He}}, \bibnamefont{and}
  \bibinfo{author}{\bibfnamefont{G.}~\bibnamefont{Li}}, \bibinfo{journal}{Nucl.
  Phys. B} \textbf{\bibinfo{volume}{953}}, \bibinfo{pages}{114968}
  (\bibinfo{year}{2020}), \eprint{1807.11363}.

\bibitem[{\citenamefont{Ilten et~al.}(2018)\citenamefont{Ilten, Soreq,
  Williams, and Xue}}]{Ilten:2018crw}
\bibinfo{author}{\bibfnamefont{P.}~\bibnamefont{Ilten}},
  \bibinfo{author}{\bibfnamefont{Y.}~\bibnamefont{Soreq}},
  \bibinfo{author}{\bibfnamefont{M.}~\bibnamefont{Williams}}, \bibnamefont{and}
  \bibinfo{author}{\bibfnamefont{W.}~\bibnamefont{Xue}},
  \bibinfo{journal}{JHEP} \textbf{\bibinfo{volume}{06}}, \bibinfo{pages}{004}
  (\bibinfo{year}{2018}), \eprint{1801.04847}.

\bibitem[{\citenamefont{Fuyuto et~al.}(2020)\citenamefont{Fuyuto, He, Li, and
  Ramsey-Musolf}}]{Fuyuto:2019vfe}
\bibinfo{author}{\bibfnamefont{K.}~\bibnamefont{Fuyuto}},
  \bibinfo{author}{\bibfnamefont{X.-G.} \bibnamefont{He}},
  \bibinfo{author}{\bibfnamefont{G.}~\bibnamefont{Li}}, \bibnamefont{and}
  \bibinfo{author}{\bibfnamefont{M.}~\bibnamefont{Ramsey-Musolf}},
  \bibinfo{journal}{Phys. Rev. D} \textbf{\bibinfo{volume}{101}},
  \bibinfo{pages}{075016} (\bibinfo{year}{2020}), \eprint{1902.10340}.

\bibitem[{\citenamefont{Du et~al.}(2020)\citenamefont{Du, Liu, and
  Tran}}]{Du:2019mlc}
\bibinfo{author}{\bibfnamefont{M.}~\bibnamefont{Du}},
  \bibinfo{author}{\bibfnamefont{Z.}~\bibnamefont{Liu}}, \bibnamefont{and}
  \bibinfo{author}{\bibfnamefont{V.~Q.} \bibnamefont{Tran}},
  \bibinfo{journal}{JHEP} \textbf{\bibinfo{volume}{05}}, \bibinfo{pages}{055}
  (\bibinfo{year}{2020}), \eprint{1912.00422}.

\bibitem[{\citenamefont{Chen et~al.}(2021)\citenamefont{Chen, Hu, Wu, and
  Yi}}]{Chen:2020bok}
\bibinfo{author}{\bibfnamefont{X.}~\bibnamefont{Chen}},
  \bibinfo{author}{\bibfnamefont{Z.}~\bibnamefont{Hu}},
  \bibinfo{author}{\bibfnamefont{Y.}~\bibnamefont{Wu}}, \bibnamefont{and}
  \bibinfo{author}{\bibfnamefont{K.}~\bibnamefont{Yi}}, \bibinfo{journal}{Phys.
  Lett. B} \textbf{\bibinfo{volume}{814}}, \bibinfo{pages}{136076}
  (\bibinfo{year}{2021}), \eprint{2001.04382}.

\bibitem[{\citenamefont{An et~al.}(2021)\citenamefont{An, Huang, Liu, and
  Xue}}]{An:2020jmf}
\bibinfo{author}{\bibfnamefont{H.}~\bibnamefont{An}},
  \bibinfo{author}{\bibfnamefont{F.~P.} \bibnamefont{Huang}},
  \bibinfo{author}{\bibfnamefont{J.}~\bibnamefont{Liu}}, \bibnamefont{and}
  \bibinfo{author}{\bibfnamefont{W.}~\bibnamefont{Xue}},
  \bibinfo{journal}{Phys. Rev. Lett.} \textbf{\bibinfo{volume}{126}},
  \bibinfo{pages}{181102} (\bibinfo{year}{2021}), \eprint{2010.15836}.

\bibitem[{\citenamefont{An et~al.}(2020)\citenamefont{An, Pospelov, Pradler,
  and Ritz}}]{An:2020bxd}
\bibinfo{author}{\bibfnamefont{H.}~\bibnamefont{An}},
  \bibinfo{author}{\bibfnamefont{M.}~\bibnamefont{Pospelov}},
  \bibinfo{author}{\bibfnamefont{J.}~\bibnamefont{Pradler}}, \bibnamefont{and}
  \bibinfo{author}{\bibfnamefont{A.}~\bibnamefont{Ritz}},
  \bibinfo{journal}{Phys. Rev. D} \textbf{\bibinfo{volume}{102}},
  \bibinfo{pages}{115022} (\bibinfo{year}{2020}), \eprint{2006.13929}.

\bibitem[{\citenamefont{Kribs et~al.}(2021)\citenamefont{Kribs, McKeen, and
  Raj}}]{Kribs:2020vyk}
\bibinfo{author}{\bibfnamefont{G.~D.} \bibnamefont{Kribs}},
  \bibinfo{author}{\bibfnamefont{D.}~\bibnamefont{McKeen}}, \bibnamefont{and}
  \bibinfo{author}{\bibfnamefont{N.}~\bibnamefont{Raj}},
  \bibinfo{journal}{Phys. Rev. Lett.} \textbf{\bibinfo{volume}{126}},
  \bibinfo{pages}{011801} (\bibinfo{year}{2021}), \eprint{2007.15655}.

\bibitem[{\citenamefont{Fabbrichesi et~al.}(2020)\citenamefont{Fabbrichesi,
  Gabrielli, and Lanfranchi}}]{Fabbrichesi:2020wbt}
\bibinfo{author}{\bibfnamefont{M.}~\bibnamefont{Fabbrichesi}},
  \bibinfo{author}{\bibfnamefont{E.}~\bibnamefont{Gabrielli}},
  \bibnamefont{and}
  \bibinfo{author}{\bibfnamefont{G.}~\bibnamefont{Lanfranchi}}
  (\bibinfo{year}{2020}), \eprint{2005.01515}.

\bibitem[{\citenamefont{Cheng et~al.}(2021)\citenamefont{Cheng, He,
  Ramsey-Musolf, and Sun}}]{Cheng:2021qbl}
\bibinfo{author}{\bibfnamefont{Y.}~\bibnamefont{Cheng}},
  \bibinfo{author}{\bibfnamefont{X.-G.} \bibnamefont{He}},
  \bibinfo{author}{\bibfnamefont{M.~J.} \bibnamefont{Ramsey-Musolf}},
  \bibnamefont{and} \bibinfo{author}{\bibfnamefont{J.}~\bibnamefont{Sun}}
  (\bibinfo{year}{2021}), \eprint{2104.11563}.

\bibitem[{\citenamefont{Dev et~al.}(2021)\citenamefont{Dev, Rodejohann, Xu, and
  Zhang}}]{Dev:2021otb}
\bibinfo{author}{\bibfnamefont{P.~S.~B.} \bibnamefont{Dev}},
  \bibinfo{author}{\bibfnamefont{W.}~\bibnamefont{Rodejohann}},
  \bibinfo{author}{\bibfnamefont{X.-J.} \bibnamefont{Xu}}, \bibnamefont{and}
  \bibinfo{author}{\bibfnamefont{Y.}~\bibnamefont{Zhang}},
  \bibinfo{journal}{JHEP} \textbf{\bibinfo{volume}{06}}, \bibinfo{pages}{039}
  (\bibinfo{year}{2021}), \eprint{2103.09067}.

\bibitem[{\citenamefont{Thomas et~al.}(2021)\citenamefont{Thomas, Wang, and
  Williams}}]{Thomas:2021lub}
\bibinfo{author}{\bibfnamefont{A.~W.} \bibnamefont{Thomas}},
  \bibinfo{author}{\bibfnamefont{X.~G.} \bibnamefont{Wang}}, \bibnamefont{and}
  \bibinfo{author}{\bibfnamefont{A.~G.} \bibnamefont{Williams}}
  (\bibinfo{year}{2021}), \eprint{2111.05664}.

\bibitem[{\citenamefont{Graham et~al.}(2021)\citenamefont{Graham, Hearty, and
  Williams}}]{Graham:2021ggy}
\bibinfo{author}{\bibfnamefont{M.}~\bibnamefont{Graham}},
  \bibinfo{author}{\bibfnamefont{C.}~\bibnamefont{Hearty}}, \bibnamefont{and}
  \bibinfo{author}{\bibfnamefont{M.}~\bibnamefont{Williams}},
  \bibinfo{journal}{Ann. Rev. Nucl. Part. Sci.} \textbf{\bibinfo{volume}{71}},
  \bibinfo{pages}{37} (\bibinfo{year}{2021}), \eprint{2104.10280}.

\bibitem[{\citenamefont{\.Inan and Kisselev}(2021)}]{Inan:2021dir}
\bibinfo{author}{\bibfnamefont{S.~C.} \bibnamefont{\.Inan}} \bibnamefont{and}
  \bibinfo{author}{\bibfnamefont{A.~V.} \bibnamefont{Kisselev}}
  (\bibinfo{year}{2021}), \eprint{2112.13070}.

\bibitem[{\citenamefont{Du et~al.}(2021)\citenamefont{Du, Fang, Liu, and
  Tran}}]{Du:2021cmt}
\bibinfo{author}{\bibfnamefont{M.}~\bibnamefont{Du}},
  \bibinfo{author}{\bibfnamefont{R.}~\bibnamefont{Fang}},
  \bibinfo{author}{\bibfnamefont{Z.}~\bibnamefont{Liu}}, \bibnamefont{and}
  \bibinfo{author}{\bibfnamefont{V.~Q.} \bibnamefont{Tran}}
  (\bibinfo{year}{2021}), \eprint{2111.15503}.

\bibitem[{\citenamefont{Barducci et~al.}(2021)\citenamefont{Barducci, Bertuzzo,
  di~Cortona, and Salla}}]{Barducci:2021egn}
\bibinfo{author}{\bibfnamefont{D.}~\bibnamefont{Barducci}},
  \bibinfo{author}{\bibfnamefont{E.}~\bibnamefont{Bertuzzo}},
  \bibinfo{author}{\bibfnamefont{G.~G.} \bibnamefont{di~Cortona}},
  \bibnamefont{and} \bibinfo{author}{\bibfnamefont{G.~M.} \bibnamefont{Salla}},
  \bibinfo{journal}{JHEP} \textbf{\bibinfo{volume}{12}}, \bibinfo{pages}{081}
  (\bibinfo{year}{2021}), \eprint{2109.04852}.

\bibitem[{\citenamefont{Su et~al.}(2021)\citenamefont{Su, Wu, and
  Zhu}}]{Su:2021jvk}
\bibinfo{author}{\bibfnamefont{L.}~\bibnamefont{Su}},
  \bibinfo{author}{\bibfnamefont{L.}~\bibnamefont{Wu}}, \bibnamefont{and}
  \bibinfo{author}{\bibfnamefont{B.}~\bibnamefont{Zhu}} (\bibinfo{year}{2021}),
  \eprint{2105.06326}.

\bibitem[{\citenamefont{Caputo et~al.}(2021)\citenamefont{Caputo, Millar,
  O'Hare, and Vitagliano}}]{Caputo:2021eaa}
\bibinfo{author}{\bibfnamefont{A.}~\bibnamefont{Caputo}},
  \bibinfo{author}{\bibfnamefont{A.~J.} \bibnamefont{Millar}},
  \bibinfo{author}{\bibfnamefont{C.~A.~J.} \bibnamefont{O'Hare}},
  \bibnamefont{and}
  \bibinfo{author}{\bibfnamefont{E.}~\bibnamefont{Vitagliano}},
  \bibinfo{journal}{Phys. Rev. D} \textbf{\bibinfo{volume}{104}},
  \bibinfo{pages}{095029} (\bibinfo{year}{2021}), \eprint{2105.04565}.

\bibitem[{\citenamefont{Wong and Huang}(2021)}]{Wong:2021lgk}
\bibinfo{author}{\bibfnamefont{X.}~\bibnamefont{Wong}} \bibnamefont{and}
  \bibinfo{author}{\bibfnamefont{Y.}~\bibnamefont{Huang}},
  \bibinfo{journal}{Eur. Phys. J. C} \textbf{\bibinfo{volume}{81}},
  \bibinfo{pages}{442} (\bibinfo{year}{2021}), \eprint{2103.15079}.

\bibitem[{\citenamefont{Thomas et~al.}(2022)\citenamefont{Thomas, Wang, and
  Williams}}]{Thomas:2022qhj}
\bibinfo{author}{\bibfnamefont{A.~W.} \bibnamefont{Thomas}},
  \bibinfo{author}{\bibfnamefont{X.~G.} \bibnamefont{Wang}}, \bibnamefont{and}
  \bibinfo{author}{\bibfnamefont{A.~G.} \bibnamefont{Williams}}
  (\bibinfo{year}{2022}), \eprint{2201.06760}.

\bibitem[{\citenamefont{Hosseini and Najafabadi}(2022)}]{Hosseini:2022urq}
\bibinfo{author}{\bibfnamefont{Y.}~\bibnamefont{Hosseini}} \bibnamefont{and}
  \bibinfo{author}{\bibfnamefont{M.~M.} \bibnamefont{Najafabadi}}
  (\bibinfo{year}{2022}), \eprint{2202.10058}.

\bibitem[{\citenamefont{Holdom}(1986)}]{Holdom:1985ag}
\bibinfo{author}{\bibfnamefont{B.}~\bibnamefont{Holdom}},
  \bibinfo{journal}{Phys. Lett. B} \textbf{\bibinfo{volume}{166}},
  \bibinfo{pages}{196} (\bibinfo{year}{1986}).

\bibitem[{\citenamefont{Batley et~al.}(2015)}]{NA482:2015wmo}
\bibinfo{author}{\bibfnamefont{J.~R.} \bibnamefont{Batley}}
  \bibnamefont{et~al.} (\bibinfo{collaboration}{NA48/2}),
  \bibinfo{journal}{Phys. Lett. B} \textbf{\bibinfo{volume}{746}},
  \bibinfo{pages}{178} (\bibinfo{year}{2015}), \eprint{1504.00607}.

\bibitem[{\citenamefont{Merkel et~al.}(2014)}]{Merkel:2014avp}
\bibinfo{author}{\bibfnamefont{H.}~\bibnamefont{Merkel}} \bibnamefont{et~al.},
  \bibinfo{journal}{Phys. Rev. Lett.} \textbf{\bibinfo{volume}{112}},
  \bibinfo{pages}{221802} (\bibinfo{year}{2014}), \eprint{1404.5502}.

\bibitem[{\citenamefont{Lees et~al.}(2014)}]{BaBar:2014zli}
\bibinfo{author}{\bibfnamefont{J.~P.} \bibnamefont{Lees}} \bibnamefont{et~al.}
  (\bibinfo{collaboration}{BaBar}), \bibinfo{journal}{Phys. Rev. Lett.}
  \textbf{\bibinfo{volume}{113}}, \bibinfo{pages}{201801}
  (\bibinfo{year}{2014}), \eprint{1406.2980}.

\bibitem[{\citenamefont{Archilli et~al.}(2012)}]{KLOE-2:2011hhj}
\bibinfo{author}{\bibfnamefont{F.}~\bibnamefont{Archilli}} \bibnamefont{et~al.}
  (\bibinfo{collaboration}{KLOE-2}), \bibinfo{journal}{Phys. Lett. B}
  \textbf{\bibinfo{volume}{706}}, \bibinfo{pages}{251} (\bibinfo{year}{2012}),
  \eprint{1110.0411}.

\bibitem[{\citenamefont{Babusci et~al.}(2013)}]{KLOE-2:2012lii}
\bibinfo{author}{\bibfnamefont{D.}~\bibnamefont{Babusci}} \bibnamefont{et~al.}
  (\bibinfo{collaboration}{KLOE-2}), \bibinfo{journal}{Phys. Lett. B}
  \textbf{\bibinfo{volume}{720}}, \bibinfo{pages}{111} (\bibinfo{year}{2013}),
  \eprint{1210.3927}.

\bibitem[{\citenamefont{Babusci et~al.}(2014)}]{KLOE-2:2014qxg}
\bibinfo{author}{\bibfnamefont{D.}~\bibnamefont{Babusci}} \bibnamefont{et~al.}
  (\bibinfo{collaboration}{KLOE-2}), \bibinfo{journal}{Phys. Lett. B}
  \textbf{\bibinfo{volume}{736}}, \bibinfo{pages}{459} (\bibinfo{year}{2014}),
  \eprint{1404.7772}.

\bibitem[{\citenamefont{Anastasi et~al.}(2016)}]{KLOE-2:2016ydq}
\bibinfo{author}{\bibfnamefont{A.}~\bibnamefont{Anastasi}} \bibnamefont{et~al.}
  (\bibinfo{collaboration}{KLOE-2}), \bibinfo{journal}{Phys. Lett. B}
  \textbf{\bibinfo{volume}{757}}, \bibinfo{pages}{356} (\bibinfo{year}{2016}),
  \eprint{1603.06086}.

\bibitem[{\citenamefont{Aaij et~al.}(2020)}]{LHCb:2019vmc}
\bibinfo{author}{\bibfnamefont{R.}~\bibnamefont{Aaij}} \bibnamefont{et~al.}
  (\bibinfo{collaboration}{LHCb}), \bibinfo{journal}{Phys. Rev. Lett.}
  \textbf{\bibinfo{volume}{124}}, \bibinfo{pages}{041801}
  (\bibinfo{year}{2020}), \eprint{1910.06926}.

\bibitem[{\citenamefont{Hook et~al.}(2011)\citenamefont{Hook, Izaguirre, and
  Wacker}}]{Hook:2010tw}
\bibinfo{author}{\bibfnamefont{A.}~\bibnamefont{Hook}},
  \bibinfo{author}{\bibfnamefont{E.}~\bibnamefont{Izaguirre}},
  \bibnamefont{and} \bibinfo{author}{\bibfnamefont{J.~G.}
  \bibnamefont{Wacker}}, \bibinfo{journal}{Adv. High Energy Phys.}
  \textbf{\bibinfo{volume}{2011}}, \bibinfo{pages}{859762}
  (\bibinfo{year}{2011}), \eprint{1006.0973}.

\bibitem[{\citenamefont{Blumlein}(2013)}]{Blumlein:2012bf}
\bibinfo{author}{\bibfnamefont{J.}~\bibnamefont{Blumlein}},
  \bibinfo{journal}{Prog. Part. Nucl. Phys.} \textbf{\bibinfo{volume}{69}},
  \bibinfo{pages}{28} (\bibinfo{year}{2013}), \eprint{1208.6087}.

\bibitem[{\citenamefont{Zyla et~al.}(2020)}]{Zyla:2020zbs}
\bibinfo{author}{\bibfnamefont{P.}~\bibnamefont{Zyla}} \bibnamefont{et~al.}
  (\bibinfo{collaboration}{Particle Data Group}), \bibinfo{journal}{PTEP}
  \textbf{\bibinfo{volume}{2020}}, \bibinfo{pages}{083C01}
  (\bibinfo{year}{2020}).

\bibitem[{\citenamefont{Cirigliano et~al.}(2021)\citenamefont{Cirigliano,
  Fuyuto, Lee, Mereghetti, and Yan}}]{Cirigliano:2021img}
\bibinfo{author}{\bibfnamefont{V.}~\bibnamefont{Cirigliano}},
  \bibinfo{author}{\bibfnamefont{K.}~\bibnamefont{Fuyuto}},
  \bibinfo{author}{\bibfnamefont{C.}~\bibnamefont{Lee}},
  \bibinfo{author}{\bibfnamefont{E.}~\bibnamefont{Mereghetti}},
  \bibnamefont{and} \bibinfo{author}{\bibfnamefont{B.}~\bibnamefont{Yan}},
  \bibinfo{journal}{JHEP} \textbf{\bibinfo{volume}{03}}, \bibinfo{pages}{256}
  (\bibinfo{year}{2021}), \eprint{2102.06176}.

\bibitem[{\citenamefont{van Neerven and Zijlstra}(1991)}]{vanNeerven:1991nn}
\bibinfo{author}{\bibfnamefont{W.~L.} \bibnamefont{van Neerven}}
  \bibnamefont{and} \bibinfo{author}{\bibfnamefont{E.~B.}
  \bibnamefont{Zijlstra}}, \bibinfo{journal}{Phys. Lett. B}
  \textbf{\bibinfo{volume}{272}}, \bibinfo{pages}{127} (\bibinfo{year}{1991}).

\bibitem[{\citenamefont{Zijlstra and van
  Neerven}(1992{\natexlab{a}})}]{Zijlstra:1992qd}
\bibinfo{author}{\bibfnamefont{E.~B.} \bibnamefont{Zijlstra}} \bibnamefont{and}
  \bibinfo{author}{\bibfnamefont{W.~L.} \bibnamefont{van Neerven}},
  \bibinfo{journal}{Nucl. Phys. B} \textbf{\bibinfo{volume}{383}},
  \bibinfo{pages}{525} (\bibinfo{year}{1992}{\natexlab{a}}).

\bibitem[{\citenamefont{Zijlstra and van
  Neerven}(1992{\natexlab{b}})}]{Zijlstra:1992kj}
\bibinfo{author}{\bibfnamefont{E.~B.} \bibnamefont{Zijlstra}} \bibnamefont{and}
  \bibinfo{author}{\bibfnamefont{W.~L.} \bibnamefont{van Neerven}},
  \bibinfo{journal}{Phys. Lett. B} \textbf{\bibinfo{volume}{297}},
  \bibinfo{pages}{377} (\bibinfo{year}{1992}{\natexlab{b}}).

\bibitem[{\citenamefont{Larin et~al.}(1997)\citenamefont{Larin, Nogueira, van
  Ritbergen, and Vermaseren}}]{Larin:1996wd}
\bibinfo{author}{\bibfnamefont{S.~A.} \bibnamefont{Larin}},
  \bibinfo{author}{\bibfnamefont{P.}~\bibnamefont{Nogueira}},
  \bibinfo{author}{\bibfnamefont{T.}~\bibnamefont{van Ritbergen}},
  \bibnamefont{and} \bibinfo{author}{\bibfnamefont{J.~A.~M.}
  \bibnamefont{Vermaseren}}, \bibinfo{journal}{Nucl. Phys. B}
  \textbf{\bibinfo{volume}{492}}, \bibinfo{pages}{338} (\bibinfo{year}{1997}),
  \eprint{hep-ph/9605317}.

\bibitem[{\citenamefont{Moch and Vermaseren}(2000)}]{Moch:1999eb}
\bibinfo{author}{\bibfnamefont{S.}~\bibnamefont{Moch}} \bibnamefont{and}
  \bibinfo{author}{\bibfnamefont{J.~A.~M.} \bibnamefont{Vermaseren}},
  \bibinfo{journal}{Nucl. Phys. B} \textbf{\bibinfo{volume}{573}},
  \bibinfo{pages}{853} (\bibinfo{year}{2000}), \eprint{hep-ph/9912355}.

\bibitem[{\citenamefont{Bierenbaum et~al.}(2007)\citenamefont{Bierenbaum,
  Blumlein, and Klein}}]{Bierenbaum:2007qe}
\bibinfo{author}{\bibfnamefont{I.}~\bibnamefont{Bierenbaum}},
  \bibinfo{author}{\bibfnamefont{J.}~\bibnamefont{Blumlein}}, \bibnamefont{and}
  \bibinfo{author}{\bibfnamefont{S.}~\bibnamefont{Klein}},
  \bibinfo{journal}{Nucl. Phys. B} \textbf{\bibinfo{volume}{780}},
  \bibinfo{pages}{40} (\bibinfo{year}{2007}), \eprint{hep-ph/0703285}.

\bibitem[{\citenamefont{Bierenbaum et~al.}(2009)\citenamefont{Bierenbaum,
  Blumlein, and Klein}}]{Bierenbaum:2009mv}
\bibinfo{author}{\bibfnamefont{I.}~\bibnamefont{Bierenbaum}},
  \bibinfo{author}{\bibfnamefont{J.}~\bibnamefont{Blumlein}}, \bibnamefont{and}
  \bibinfo{author}{\bibfnamefont{S.}~\bibnamefont{Klein}},
  \bibinfo{journal}{Nucl. Phys. B} \textbf{\bibinfo{volume}{820}},
  \bibinfo{pages}{417} (\bibinfo{year}{2009}), \eprint{0904.3563}.

\bibitem[{\citenamefont{Guzzi et~al.}(2012)\citenamefont{Guzzi, Nadolsky, Lai,
  and Yuan}}]{Guzzi:2011ew}
\bibinfo{author}{\bibfnamefont{M.}~\bibnamefont{Guzzi}},
  \bibinfo{author}{\bibfnamefont{P.~M.} \bibnamefont{Nadolsky}},
  \bibinfo{author}{\bibfnamefont{H.-L.} \bibnamefont{Lai}}, \bibnamefont{and}
  \bibinfo{author}{\bibfnamefont{C.~P.} \bibnamefont{Yuan}},
  \bibinfo{journal}{Phys. Rev. D} \textbf{\bibinfo{volume}{86}},
  \bibinfo{pages}{053005} (\bibinfo{year}{2012}), \eprint{1108.5112}.

\bibitem[{\citenamefont{Kawamura et~al.}(2012)\citenamefont{Kawamura,
  Lo~Presti, Moch, and Vogt}}]{Kawamura:2012cr}
\bibinfo{author}{\bibfnamefont{H.}~\bibnamefont{Kawamura}},
  \bibinfo{author}{\bibfnamefont{N.~A.} \bibnamefont{Lo~Presti}},
  \bibinfo{author}{\bibfnamefont{S.}~\bibnamefont{Moch}}, \bibnamefont{and}
  \bibinfo{author}{\bibfnamefont{A.}~\bibnamefont{Vogt}},
  \bibinfo{journal}{Nucl. Phys. B} \textbf{\bibinfo{volume}{864}},
  \bibinfo{pages}{399} (\bibinfo{year}{2012}), \eprint{1205.5727}.

\bibitem[{\citenamefont{Kang et~al.}(2014)\citenamefont{Kang, Lee, and
  Stewart}}]{Kang:2014qba}
\bibinfo{author}{\bibfnamefont{D.}~\bibnamefont{Kang}},
  \bibinfo{author}{\bibfnamefont{C.}~\bibnamefont{Lee}}, \bibnamefont{and}
  \bibinfo{author}{\bibfnamefont{I.~W.} \bibnamefont{Stewart}},
  \bibinfo{journal}{JHEP} \textbf{\bibinfo{volume}{11}}, \bibinfo{pages}{132}
  (\bibinfo{year}{2014}), \eprint{1407.6706}.

\bibitem[{\citenamefont{Aaron et~al.}(2012)}]{H1:2012qti}
\bibinfo{author}{\bibfnamefont{F.~D.} \bibnamefont{Aaron}} \bibnamefont{et~al.}
  (\bibinfo{collaboration}{H1}), \bibinfo{journal}{JHEP}
  \textbf{\bibinfo{volume}{09}}, \bibinfo{pages}{061} (\bibinfo{year}{2012}),
  \eprint{1206.7007}.

\bibitem[{\citenamefont{Chekanov et~al.}(2009)}]{ZEUS:2009swh}
\bibinfo{author}{\bibfnamefont{S.}~\bibnamefont{Chekanov}} \bibnamefont{et~al.}
  (\bibinfo{collaboration}{ZEUS}), \bibinfo{journal}{Eur. Phys. J. C}
  \textbf{\bibinfo{volume}{62}}, \bibinfo{pages}{625} (\bibinfo{year}{2009}),
  \eprint{0901.2385}.

\bibitem[{\citenamefont{Abramowicz et~al.}(2013)}]{ZEUS:2012zcp}
\bibinfo{author}{\bibfnamefont{H.}~\bibnamefont{Abramowicz}}
  \bibnamefont{et~al.} (\bibinfo{collaboration}{ZEUS}), \bibinfo{journal}{Phys.
  Rev. D} \textbf{\bibinfo{volume}{87}}, \bibinfo{pages}{052014}
  (\bibinfo{year}{2013}), \eprint{1208.6138}.

\bibitem[{\citenamefont{Hou et~al.}(2021)}]{Hou:2019efy}
\bibinfo{author}{\bibfnamefont{T.-J.} \bibnamefont{Hou}} \bibnamefont{et~al.},
  \bibinfo{journal}{Phys. Rev. D} \textbf{\bibinfo{volume}{103}},
  \bibinfo{pages}{014013} (\bibinfo{year}{2021}), \eprint{1912.10053}.

\bibitem[{\citenamefont{Yan et~al.}(2021)\citenamefont{Yan, Yu, and
  Yuan}}]{Yan:2021htf}
\bibinfo{author}{\bibfnamefont{B.}~\bibnamefont{Yan}},
  \bibinfo{author}{\bibfnamefont{Z.}~\bibnamefont{Yu}}, \bibnamefont{and}
  \bibinfo{author}{\bibfnamefont{C.~P.} \bibnamefont{Yuan}},
  \bibinfo{journal}{Phys. Lett. B} \textbf{\bibinfo{volume}{822}},
  \bibinfo{pages}{136697} (\bibinfo{year}{2021}), \eprint{2107.02134}.

\bibitem[{\citenamefont{Li et~al.}(2021)\citenamefont{Li, Yan, and
  Yuan}}]{Li:2021uww}
\bibinfo{author}{\bibfnamefont{H.~T.} \bibnamefont{Li}},
  \bibinfo{author}{\bibfnamefont{B.}~\bibnamefont{Yan}}, \bibnamefont{and}
  \bibinfo{author}{\bibfnamefont{C.~P.} \bibnamefont{Yuan}}
  (\bibinfo{year}{2021}), \eprint{2112.07747}.

\bibitem[{\citenamefont{Abdul~Khalek et~al.}(2021)}]{AbdulKhalek:2021gbh}
\bibinfo{author}{\bibfnamefont{R.}~\bibnamefont{Abdul~Khalek}}
  \bibnamefont{et~al.} (\bibinfo{year}{2021}), \eprint{2103.05419}.

\bibitem[{\citenamefont{Arcadi et~al.}(2020)\citenamefont{Arcadi, Lindner,
  Martins, and Queiroz}}]{Arcadi:2019uif}
\bibinfo{author}{\bibfnamefont{G.}~\bibnamefont{Arcadi}},
  \bibinfo{author}{\bibfnamefont{M.}~\bibnamefont{Lindner}},
  \bibinfo{author}{\bibfnamefont{J.}~\bibnamefont{Martins}}, \bibnamefont{and}
  \bibinfo{author}{\bibfnamefont{F.~S.} \bibnamefont{Queiroz}},
  \bibinfo{journal}{Nucl. Phys. B} \textbf{\bibinfo{volume}{959}},
  \bibinfo{pages}{115158} (\bibinfo{year}{2020}), \eprint{1906.04755}.

\bibitem[{\citenamefont{Bilmis et~al.}(2015)\citenamefont{Bilmis, Turan, Aliev,
  Deniz, Singh, and Wong}}]{Bilmis:2015lja}
\bibinfo{author}{\bibfnamefont{S.}~\bibnamefont{Bilmis}},
  \bibinfo{author}{\bibfnamefont{I.}~\bibnamefont{Turan}},
  \bibinfo{author}{\bibfnamefont{T.~M.} \bibnamefont{Aliev}},
  \bibinfo{author}{\bibfnamefont{M.}~\bibnamefont{Deniz}},
  \bibinfo{author}{\bibfnamefont{L.}~\bibnamefont{Singh}}, \bibnamefont{and}
  \bibinfo{author}{\bibfnamefont{H.~T.} \bibnamefont{Wong}},
  \bibinfo{journal}{Phys. Rev. D} \textbf{\bibinfo{volume}{92}},
  \bibinfo{pages}{033009} (\bibinfo{year}{2015}), \eprint{1502.07763}.

\end{thebibliography}

\end{document}